\RequirePackage[hyphens]{url} %% allows wordwrap of long urls, needs to be before document class
\documentclass{soups} 
\usepackage{times}
\usepackage{balance}
\usepackage{float}
\usepackage{xcolor}
\usepackage{wrapfig}
\usepackage{graphicx}
\graphicspath{ {images/} }
\usepackage{lipsum}
\usepackage{url}
\usepackage{hyperref}

\usepackage{subcaption}
\usepackage{enumitem} %% allows specification of the tab size in enumerate, itemize, ... environments 

\usepackage[compress]{cite}

\begin{document}
%
% --- Author Metadata here ---
\conferenceinfo{USENIX Symposium on Usable Privacy and Security
  (SOUPS)}{2018. \\ August~12--14,~2018, Baltimore,~MD,~USA. }
\CopyrightYear{2018} 

% --- End of Author Metadata ---

\title{Forming IDEAS \\
Interactive Data Exploration \& Analysis System
\titlenote{\scriptsize{This manuscript has been authored by UT-Battelle, LLC under Contract No. DE-AC05-00OR22725 with the U.S. Department of Energy.  The United States Government retains and the publisher, by accepting the article for publication, acknowledges that the United States Government retains a non-exclusive, paid-up, irrevocable, world-wide license to publish or reproduce the published form of this manuscript, or allow others to do so, for United States Government purposes.  The Department of Energy will provide public access to these results of federally sponsored research in accordance with the DOE Public Access Plan \url{http://energy.gov/downloads/doe-public-access-plan}.}}
}
\subtitle{Configurable Visual Analytics for Cyber Security Analysts}

\author{
\alignauthor
    Robert Bridges, Maria Vincent, Kelly Huffer, John Goodall, Jessie Jamieson\textsuperscript{\dag}, Zachary Burch\textsuperscript{\ddag}\\
    \affaddr{Oak Ridge National Laboratory, Oak Ridge, TN, \textsuperscript{\dag}University Nebraska, Lincoln, NE},\\  \textsuperscript{\ddag}Virginia Tech, Blacksburg, VA
    \email{\{bridgesra, vincentms, hufferkm, jgoodall\}@ornl.gov, jdjamieson@huskers.unl.edu, zchryb@vt.edu}
}

\maketitle

\begin{abstract}
Modern cyber security operations collect an enormous amount of logging and alerting data. 
While analysts have the ability to query and compute simple statistics and plots from their data, current analytical tools are too simple to admit deep understanding. 
To detect advanced and novel attacks, analysts turn to manual investigations.
While commonplace, current investigations are time-consuming, intuition-based, and proving insufficient. 
Our hypothesis is that arming the analyst with easy-to-use data science tools will increase their work efficiency, provide them with the ability to resolve hypotheses with scientific inquiry of their data, and support their decisions with evidence over intuition. 
To this end, we present our work to build IDEAS (Interactive Data Exploration and Analysis System).
We present three real-world use cases that drive the system design from the algorithmic capabilities to the user interface. 
Finally, a modular and scalable software architecture is discussed along with plans for our pilot deployment with a security operation center. 
\end{abstract}

\section{Introduction}
\label{sec:intro}
 
Cyber operations\textemdash comprised of security analysts and charged with keeping enterprise networks secure and healthy\textemdash now have widespread data collection and query capabilities, as well as many automated alerting tools. 
Heterogeneous streams of logs are generated by firewalls, network- and host-level intrusion detection and prevention systems, packet captures, DNS and other servers, and workstations. 
While this valuable sea of data is collected and stored for analysts use, leveraging it effectively is a current challenge.

We interviewed analysts from six security centers and  observed that it is now standard to have widespread log collection and query capabilities  using Splunk (\url{www.splunk.com}) and Elasticsearch (\url{www.elastic.co}) as common log management tools. 
These tools support basic analytics; e.g., Splunk integrates with Tableau (\url{www.tableau.com})  and Elasticsearch with Kirbana facilitating fast visualization and computation of many basic tables, plots, and statistics. 

Yet, incident response and more general protection against advanced attacks relies mainly on intuition and tedious manual investigations of the relevant charts and logs.  
As an example shared with us, upon a ransomware infection, obtaining the past host logs from infected machines is a quick query, but understanding the malware behavior by analyzing logs was reported as a tedious manual process taking $\sim$80 man hours.  
In another interview an analyst described writing one-off code to parse data to hunt for anomalous lateral movement (connections to within-network IPs); the analyst wished to cluster hosts by the ratio of their DNS queries to the number of times their domain was queried and then explore anomalies over different time scales, but was unable to produce the results in a reasonable amount of time. 
In summary, analysts require scientific inquiry of their data to resolve hypotheses.  
Next generation security requires flexible machine learning tools allowing analysts to leverage their expertise and data to make decisions backed by evidence. %environment knowledge to enable scientific enquiry of their data. 
% \begin{figure}
\begin{wrapfigure}[18]{l}{0pt}
\vspace{-.34cm}
% \hspace{-.1cm}
\includegraphics[width = .23\textwidth]{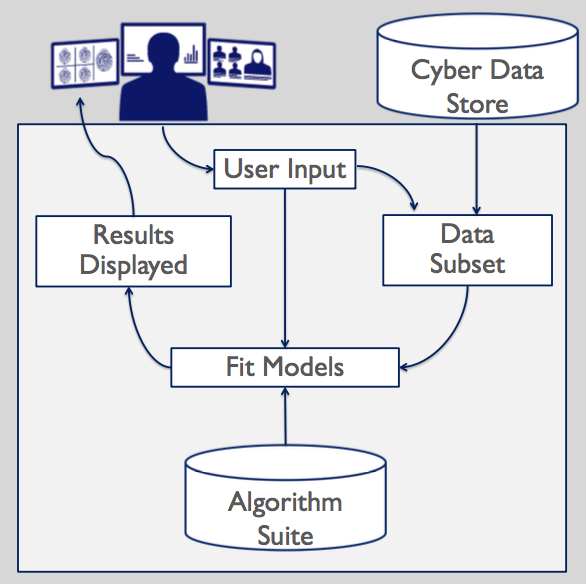}
\vspace{-.22cm}
\caption{IDEAS connects to cyber data store. Analysts can select algorithmic operations to run on subsets of their data and receive interactive visualizations of the results.}
\label{fig:system}
\end{wrapfigure}

% \end{figure}
In response, we present IDEAS, a novel, visual analytic platform designed to bring data science to the security worker. 
After connecting to the cyber operations' data store, IDEAS will give analysts options for configuring a few predefined algorithms on their data\textemdash they can choose a data subset and an analytic operation to be executed on that subset\textemdash and presents the analysts with the operation's results via an interactive visualization. 
Algorithmic implementation and model fitting  will occur ``under the hood'', requiring some user training but no algorithmic expertise. 
See Figure~\ref{fig:system}.

We met with analysts regularly at the beginning of the project to understand their data, roles, and work flows, and then discussed specific problems they encountered or capabilities they would benefit from having.
As a result of these discussions, we isolated three use-cases from real investigations that drove our algorithmic and visualization design.
These use-cases are presented as particular examples illustrating potential uses of the system. 
Algorithms were enriched or tweaked to provide understandable results, and novel visualizations crafted for both inputs and outputs of the system. 
Actual historical data was shared with us for this development, and we corresponded with operators as we progressed. 
We expect IDEAS to be sufficiently developed in the next few months to allow a pilot deployment with a security operation center, and expect iteration with them for refining the system. 
We discuss our software architectural design, highlighting benefits for deployment and scalability.

We envision many benefits of IDEAS. 
Currently-manual investigations will be semi-automated, saving valuable analyst time.  
Analysts can test and evolve hypotheses using a more scientific process; hence, evidence from results of the analytics will inform decisions over only intuition. 
Useful analytic configurations of the system can be shared without sharing (generally sensitive) data. 
This will enable intra- and inter-organizational intelligence sharing; e.g., junior analysts can execute pre-configured operations that were discovered only by insight of experienced analysts.

\section{Algorithms \& User Interface} 
\label{sec:algorithms}

\begin{figure}
% \begin{wrapfigure}[19]{l}{.57\linewidth}
\vspace{-.4cm}
\hspace{-.45cm}
\includegraphics[width = .52\textwidth]{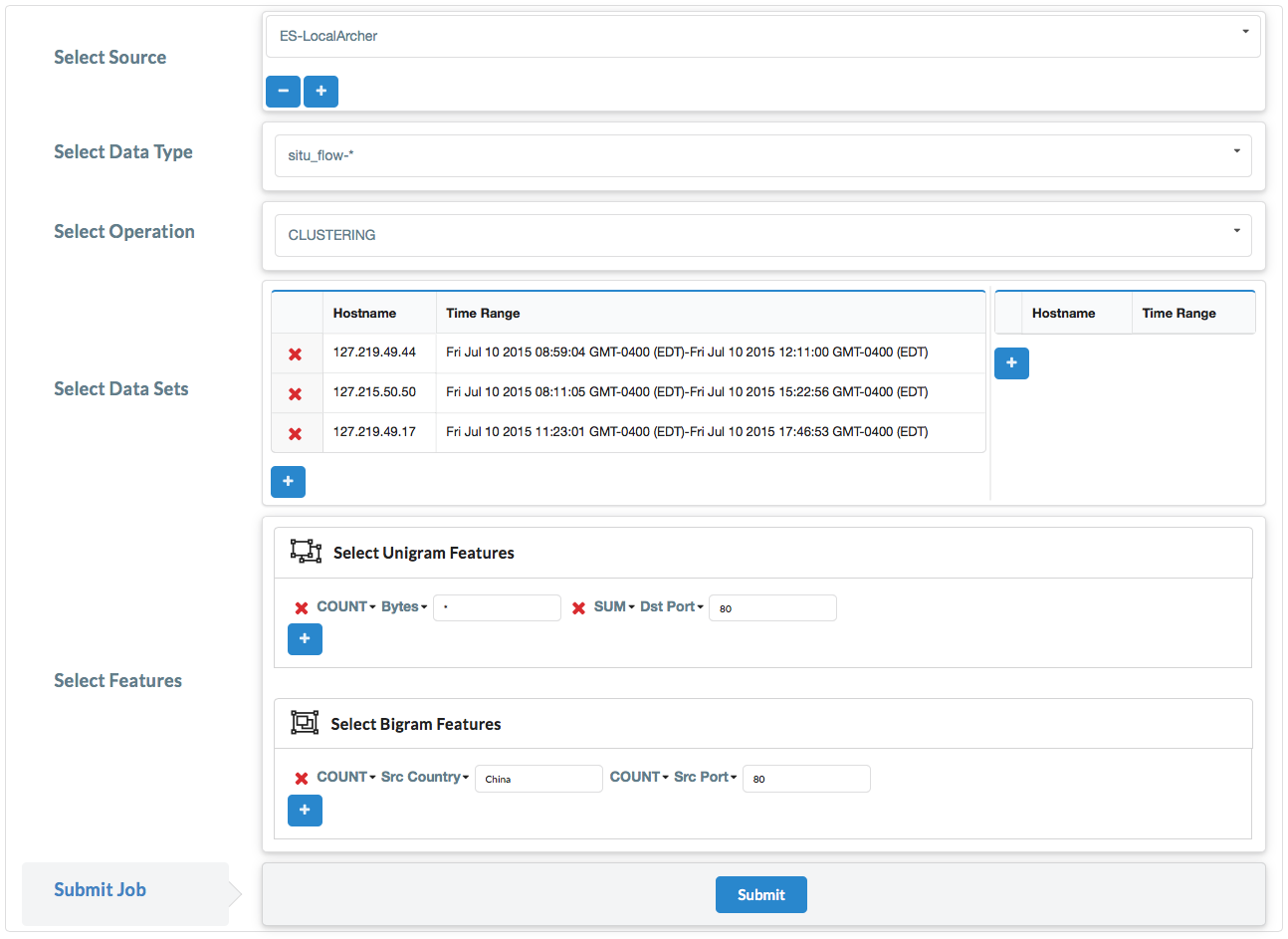}
\vspace{-.5cm}
\caption{IDEAS Input Page}
\label{fig:system-input}
% \end{wrapfigure}
\end{figure}

Informed by interactions with operators, we chose three cases disclosed by operators for the initial system implementation. 
Importantly, users only need to know the conceptual capability of each operation, not algorithmic or implementation details. 
To facilitate this, we researched the appropriate classes of algorithms and chose a fixed algorithm that has the desired operational properties. 

To configure IDEAS, users must specify two things: input dataset(s) and an operation. 
We have designed and implemented a custom user interface (UI) with a set form\textemdash a sequence of selections, drop-down boxes, and other interactive interfaces\textemdash for specifying inputs. 
All anticipated cyber data can be indexed by IP address and time; hence, datasets are chosen by indicating the type of data (network flow, host logs, etc.), and the desired [IP, time-window] tuples. 
Since algorithms require data represented as feature vectors, i.e., a collection of attribute values, the fields of the chosen dataset are displayed and can be chosen by the user. 
We are adding functionality to include combinations of fields  and simple statistics (e.g, sum of a field's values over IPs/times) as attributes. See Figure~\ref{fig:system-input}. 

Once an operation and dataset is chosen through the input UI, the system queries the cyber data store for the selected data, parses it into vectors (lists of the user-selected attributes), runs the desired algorithm, and displays the results. 
A ``jobs'' page is displayed to the users, itemizing all previous configurations and links to the results. 
The rest of this section describes each of the three operations and the corresponding output visualizations designed.

\begin{figure}
%\begin{wrapfigure}[19]{l}{.57\linewidth}
\vspace{-.4cm}
% \hspace{-.45cm}
\includegraphics[width = .5\textwidth]{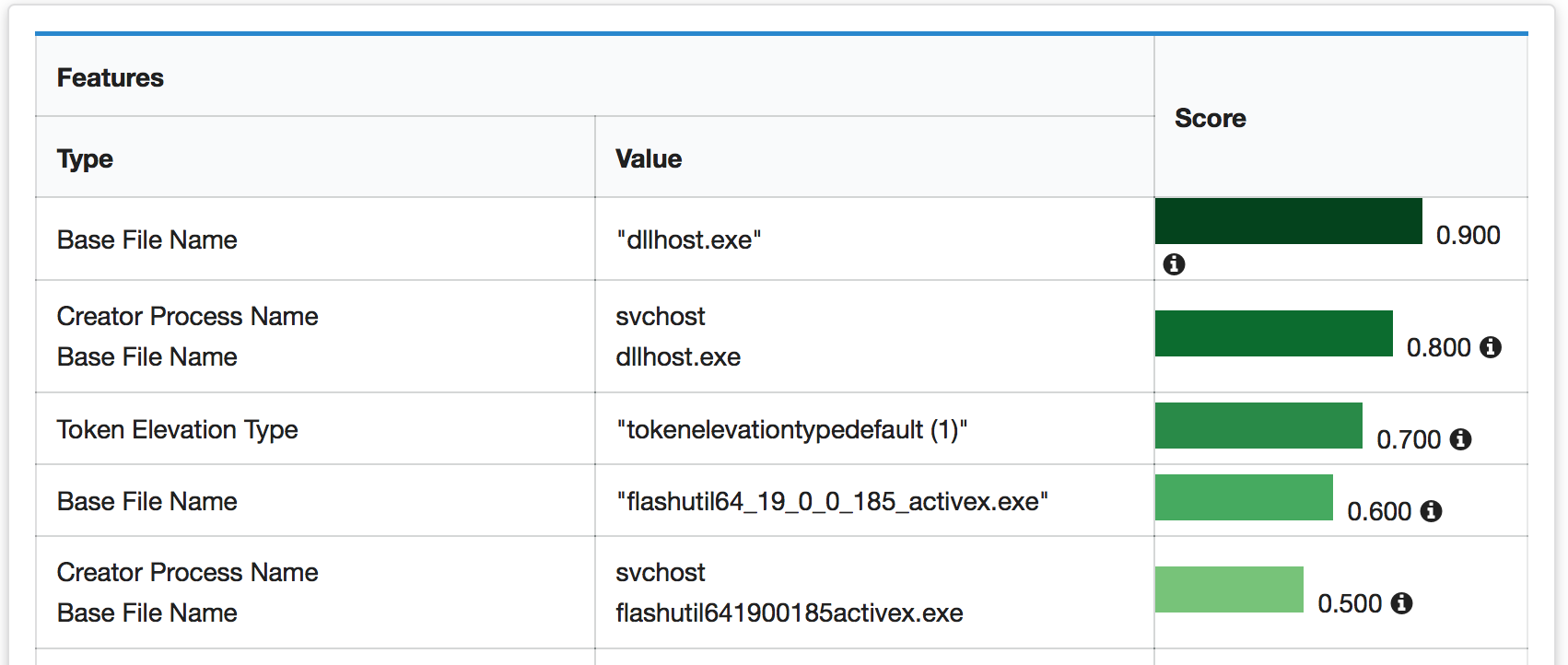}
\vspace{-.5cm}
\caption{IDEAS' Discriminant Analysis Results}
\label{fig:lda-vis}
% \end{wrapfigure}
\end{figure}

{\bf Discriminant Analysis:} Conceptually, this operation takes two datasets as input and identifies those attributes that distinguish the first set from the second. 
Algorithmically, we implement Fisher's Linear Discriminant Analysis~\cite{welling2005fisher}, a binary classifier trained on the two input sets to extract the direction in the attribute space that best separates the two sets while keeping each set tightly clustered. 
From this vector we can present to the user a list of the attributes ranked and scored by their discrimination power. 
The first few results from a particular dataset involving Cryptowall ransomware are shown in Figure~\ref{fig:lda-vis}, which depicts IDEAS' visualization produced by running our discriminant analysis algorithm. 
This example indicates that the ``Base File Name'' attribute with a value of ``dlhost.exe'' is the most discriminating attribute from this particular dataset. 
The results shown in Figure~\ref{fig:lda-vis} align with the actions actually observed in the host logs infected with Cryptowall ransomware. See Section~\ref{sec:use-cases} for more details on this use case.

{\bf Anomaly Scoring:} For the anomaly scoring operation, users specify a baseline dataset and a test dataset. For each element of the test set the algorithm furnishes a score of anomalousness relative to the baseline. 
Because datasets naturally have different distributions, we seek an anomaly score that is easy to understand and comparable across distributions. 
Following our previous work~\cite{ferragut2012new, bridges2017setting, goodall2018situ}, we estimate a distribution $f$ and compute the p-value of each data point: $pv(x_0) = \sum_{A} f(x)$ with the sum over $A:={\{x: f(x)\leq f(x_0)\}}$. 
This formula gives the percent of the distribution that is less likely than $x_0$, just like a percentile is comparable across distributions. 
To the user we present the anomaly score 1-$pv \in [0,1]$, so high scores are anomalous. 
The visualization is similar to Figure~\ref{fig:lda-vis}, but attribute values are ranked and displayed with anomaly scores. 
Matching our clustering algorithm, we use a $k-$nearest neighbor (KNN) density estimate (fit to the baseline set).

{\bf Clustering:} This operation accepts a single dataset and returns a partition into clusters (subsets that are similar) and outliers (data points that are unlike any found cluster). 
We chose HDBSCAN~\cite{campello2013density}, a hierarchical density-based clustering algorithm, because it has many advantages: it is user-friendly, requiring a single, optional parameter (research has empirically showed robustness of this parameter), and importantly, not requiring users to specify the number of clusters; it has a straightforward implementation that is fast and scalable~\cite{mcinnes2017hdbscan}; it can accommodate non-convex clusters (unlike k-means and variance-based algorithms) and clusters with a variety of densities (unlike DBSCAN~\cite{ester1996density}); it automatically identifies outliers and it provides a measure of cluster stability, telling how tightly packed the cluster is. 
An added advantage for our setting is the algorithm is based on a KNN-estimated probability distribution, allowing us to enrich the output by annotating all data points with anomaly scores using the same p-value method discussed above. 

\begin{figure}
% \begin{wrapfigure}[19]{l}{.57\linewidth}
\vspace{-.4cm}
\hspace{-.45cm}
\includegraphics[width = .5\textwidth]{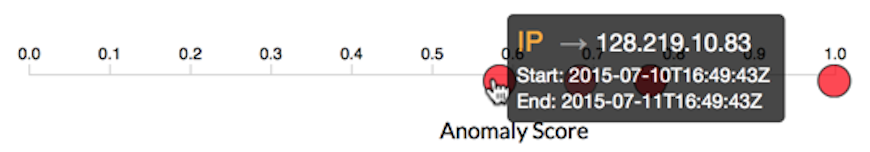}
\vspace{-.3cm}
\caption{IDEAS Outliers Visualization.}
\label{fig:outliers-vis}
% \end{wrapfigure}
\end{figure}

\begin{figure}
% \vspace{1cm}
% \begin{wrapfigure}[19]{l}{.57\linewidth}
\vspace{-.3cm}
\hspace{-.45cm}
\includegraphics[width = .5\textwidth]{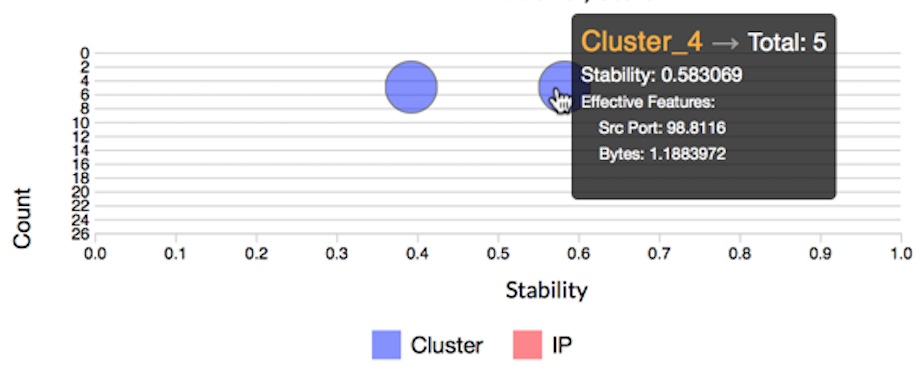}
\vspace{-.4cm}
\caption{IDEAS Clusters Visualization.}
\label{fig:clusters-vis}
% \end{wrapfigure}
\end{figure}

The clustering output required a custom UI, as current visualizations for HDBSCAN do not scale and do not accommodate many desired properties for our setting. 
IDEAS' clustering visualization consists of two complimentary charts. Figure~\ref{fig:outliers-vis} displays a linear chart with identified outliers (each red circle is a single data point not belonging to a cluster) and their placement on the line corresponds to their calculated anomaly scores and additional identifying metadata. 
That is, anomaly scoring, as computed using the algorithm above, is computed automatically for outliers identified by the clustering algorithm and displayed. 
Figure~\ref{fig:clusters-vis} visualizes the clusters with their stability (a measure of the relative density of the cluster) on X-axis and the count of data points within the cluster on the Y-axis, which is also  indicated by cluster radius.
To enhance this cluster visualization and enable result exploration, the user has the ability to break clusters into their sub-clusters and walk up and down through a cluster  tree hierarchy inherent to the HDBSCAN algorithm. 

We anticipate users will need to know what defines each cluster in terms of the input set's attributes. 
To facilitate this understanding, we run our discriminant analysis algorithm in a one-versus-all scheme, to compare each cluster to all other data points. 
This allows us to annotate each cluster with its discriminating attributes, and these are displayed in the visualization along with additional metadata, such as cluster size and stability.

\section{Operational Use Cases} 
\label{sec:use-cases}
Our workflow for research was to interview analysts and discuss problematic investigations or gaps in their current capabilities. 
This focused our developments and provided example configurations for our system. 

{\bf Automated Malware Forensics:} 
The driving example for the discriminant analysis operation is a historical investigation of a Cryptowall 3.0 ransomware outbreak at a large organization, where analysts spent approximately 80 man hours manually comparing infected host logs to ambient logs in search of the footprint left by the malware before encryption of user data. 
To execute this example using our system, the user would first choose the discriminant analysis operation, then specify the host logs of infected IPs during an interval up to encryption as the first dataset, and select a collection of similar sized slices of host logs from uninfected IPs as the other dataset. 
Lastly, the operator would choose attributes of the host logs they believe are influenced by ransomware (e.g., new files' names, extensions, and creator processes). 
The system will query their cyber data store for the host logs, parse the data into lists of attributes (e.g., counts of each new filename, extension, creator process), and  execute the algorithm. 
The output display will show each attribute instance with a score of how well it distinguishes the first (infected) data from the second (uninfected) data. 
Ideally, the ability to flexibly configure a collection of IDEAS operations as above would allow analysts to perform the forensic effort in a few hours. 

To verify the validity of our approach, we obtained actual logs from a few Windows hosts that were infected with then-zero-day ransomware Cryptowall 3.0\footnote{\noindent \url{https://digitalguardian.com/blog/detecting-cryptowall-3-using-real-time-event-correlation},  \url{https://sentinelone.com/blogs/anatomy-of-cryptowall-3-0-a-look-inside-ransomwares-tactics/}.} and uninfected hosts' logs. 
We extracted attributes by counting what file names, extensions, executed command line strings, and creator processes. 
The first few results of the IDEAS linear discriminant analysis are shown in Figure~\ref{fig:lda-vis}. 
%This example indicates that the ``Base File Name'' attribute with a value of ``dlhost.exe'' and Flash player both being run with the creator process of ``svchost''  are some of the most discriminating attributes in this particular dataset. 
We verified the extracted attributes of Cryptowall actions with analysts involved in the manual investigation of logs; further, many of the CryptoWall actions discussed in the footnoted URLs were identified as top ranking attributes by IDEAS. 
See our previous work for detailed experiments testing this concept using WannaCry Ransomware~\cite{chen2017automated} and other discriminant analysis methods. 

{\bf Situational Awareness of IP's Roles:} Analysts expressed a blind spot in their situational awareness, namely, not understanding the network roles of machines; e.g., a workstation may be configured as a web server, but without this knowledge analysts could mistake the large amount of traffic to that host for an attack. 
We build on our previous work~\cite{huffer2017sansr} to provide the insight through clustering. 
Using our system, analysts would select the IPs and time windows desired, and choose the percent of traffic on each well-known port (0-1023) in each direction. 
As an example of information learned from the output,  web servers form  two clusters with discriminating attribute being most traffic is inbound on port 80 (http) and 443 (https), while most workstations form two clusters as  web clients, with most data outbound to the same two ports.

{\bf Hunting Advanced Persistent Threats:} Advanced Persistent Threats (APTs) are adversaries using advanced/novel techniques to maintain a long-term, stealth foothold in the network. 
A general technique for identifying an APT used by security operations is to look for lateral movement; e.g., some operators write custom code to plot graph statistics of intra-organization IP-IP communications and look at the changes over time. 
We propose using the anomaly detection operation of our system to expedite this process.  
Specifically, using DNS data or network flow data, operators could baseline the number of internal IPs a given host contacts and/or the specific internal IPs contacted per day and produce an anomaly score for each subsequent day. 
This will effectively automate the custom code, and provide a principled framework to identify anomalies. 
Unlike the other two use cases, we have not yet experimentally tested this configuration, but are working towards validating this concept.

\section{Software Architecture} 
\label{sec:soft-arch} 
\begin{figure}
% \begin{wrapfigure}[19]{l}{.57\linewidth}
\vspace{-.6cm}
\hspace{-.45cm}
\includegraphics[width = .55\textwidth]{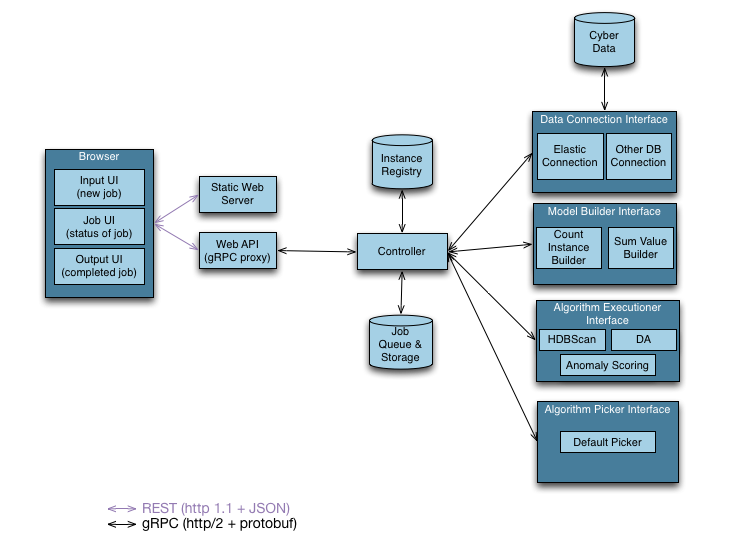}
\vspace{-.5cm}
\caption{IDEAS software architecture depicted.}
\label{fig:soft-arch}
% \end{wrapfigure}
\end{figure}

To provide a modular and flexible system, we developed a microservices architecture.  
To this end, we use Google's Protocol Buffers~\cite{varda2008protocol, protobufs}, which provide ``a language-neutral, platform-neutral, extensible way of serializing structured data'', and transmit the data using gRPC (\url{https://grpc.io})%~\cite{gRPC}
, a framework for remote procedure calls. 
Figure~\ref{fig:soft-arch} depicts IDEAS' current software architechture. 
This defines what components are needed, the topology of how they communicate, and definitions for what data types they input/output.  
The benefits of this design include the following: each component can facilitate language-agnostic code for processing and still communicate in a standard way; new components can be easily incorporated; modularity by design will facilitate changes in the architecture when needed; use of HTTP/2, a faster, more secure protocol than HTTP. 
To facilitate easy deployment, we have wrapped the system with Docker~\cite{merkel2014docker}, an open-source, virtualized container.   
The system is designed to interface with Elasticsearch%~\cite{gormley2015elasticsearch}
, as this is an increasingly popular datastore for cyber operations.

\section{Conclusion \& Plans}
This paper introduces IDEAS, an interactive visual platform designed to let cyber security analysts flexibly configure standard data science algorithms on their cyber data. 
Large cyber operations now have the capability to continually collect and query an enormous amount of data, but they rely mainly on manual investigations and very simple analytics. 
IDEAS seeks to give analysts the capability to leverage both their expertise and  powerful data science tools with no algorithmic expertise. 
Working with actual analysts, we targeted the design of IDEAS around problematic incidents or gaps in their capabilities. 
Three real-world use cases are presented to illustrate the benefit of IDEAS, once sufficiently developed. 
A high-level overview of software engineering design choices is given.

IDEAS is prototyped with limited functionality in each component, and future work to expand the capability is due. 
We are on the verge of an initial deployment with a large cyber operation for testing and refinement. 
The system currently can accept network flow data and will be extended to host logs and DNS data next. 
Currently, all three operations are implemented and current work is focusing  on the input UI to admit a wider variety of attributes. 
% Of course, future research to continually expand the library of operations is expected. 

\section{Acknowledgments}
Research sponsored by the Laboratory Directed Research and Development Program of Oak Ridge National Laboratory, managed by UT-Battelle, LLC, for the U. S. Department of Energy. Jessie D. Jamieson was supported by National Science Foundation Graduate Research Fellowship under Grant No. 25-0517-0143-002.
Any conclusions or recommendations expressed in this material are those of the authors(s) and do not necessarily reflect the views of the National Science Foundation.

\balance  %balance the two columns at the end of the paper

\small
\bibliographystyle{abbrv}
\bibliography{refs}  
\end{document}